\documentclass[lettersize,journal]{IEEEtran}
\usepackage{amsmath,amsfonts,amssymb}
\usepackage[caption=false,font=normalsize,labelfont=sf,textfont=sf]{subfig}
\usepackage{graphicx}
\usepackage{cite}
\usepackage{xcolor}
\usepackage{tikz}
\usepackage{hyperref}
\usepackage[capitalise,nameinlink]{cleveref}

\Crefname{equation}{Eq.}{Eqs.}
\Crefname{figure}{Fig.}{Figs.}
\Crefname{tabular}{Table}{Tables}
\Crefname{section}{Section}{Sections}

\def\BibTeX{{\rm B\kern-.05em{\sc i\kern-.025em b}\kern-.08em
    T\kern-.1667em\lower.7ex\hbox{E}\kern-.125emX}}

\definecolor{lime}{HTML}{A6CE39}
\DeclareRobustCommand{\orcidicon}{%
	\begin{tikzpicture}
	\draw[lime, fill=lime] (0,0) 
	circle [radius=0.16] 
	node[white] {{\fontfamily{qag}\selectfont \tiny ID}};
	\draw[white, fill=white] (-0.0625,0.095) 
	circle [radius=0.007];
	\end{tikzpicture}
	\hspace{-2mm}
}

\foreach \x in {A, ..., Z}{%
	\expandafter\xdef\csname orcid\x\endcsname{\noexpand\href{https://orcid.org/\csname orcidauthor\x\endcsname}{\noexpand\orcidicon}}
}

\makeatletter
\def\ps@IEEEtitlepagestyle{%
  \def\@oddfoot{\mycopyrightnotice}%
  \def\@oddhead{\hbox{}\@IEEEheaderstyle\leftmark\hfil\thepage}\relax
  \def\@evenhead{\@IEEEheaderstyle\thepage\hfil\leftmark\hbox{}}\relax
  \def\@evenfoot{}%
}
\def\mycopyrightnotice{%
  \begin{minipage}{\textwidth}
  \centering \scriptsize
  This work has been submitted to the IEEE for possible publication. Copyright may be transferred without notice, after which this version may no longer be accessible.
  \end{minipage}
}
\makeatother

\begin{document}
\title{Effects of Linear Modulation of Electrotactile Signals Using a Novel Device on Sensation Naturalness and Perceptual Intensity}

\author{
Amirhossein~Bayat\textsuperscript{§}\orcidA, Melika~Emami\textsuperscript{§}\orcidB, Rahim~Tafazolli\orcidC,~\IEEEmembership{Senior Member,~IEEE,}
and Atta~Quddus\orcidD%
\thanks{\textsuperscript{§}Equal contribution.}%
\thanks{All authors are with the Institute for Communication Systems (ICS), Home of 5G \& 6G Innovation Centres, University of Surrey, UK.}
}

\maketitle

\bstctlcite{IEEEexample:BSTcontrol}

\begin{abstract}
Electrotactile feedback is a promising method for delivering haptic sensations, but challenges such as the quality and naturalness of sensations hinder its adoption in commercial devices. In this study, we introduce a novel electrotactile device that enables the exploration of more complex stimulation signals to enhance sensation naturalness. We designed six stimulation signals with linearly modulated frequency, amplitude, or both, across two frequency levels based on a ramp-and-hold shape, aiming to replicate the sensation of pressing a button. Our results showed that these modulated signals achieve higher naturalness scores than tonic stimulations, with a 6.8\% improvement.
Additionally, we examined the relationship between perceived intensity and signal energy for these complex stimulation patterns. Our findings indicate that, under conditions of constant perceived intensity, signal energy is not uniform across different stimulation patterns. Instead, there is a distinct relationship between the energy levels of different patterns, which is consistently reflected in the energy of the stimulations selected by the participants. Based on our findings, we propose a predictive model that estimates the desired intensity for any stimulation pattern using this relationship between signal energies and the user’s preferred intensity for a single reference pattern. This model demonstrated high reliability, achieving a mean R$^2$ score of 83.33\%. Using this approach, intensity calibration for different stimulation patterns can be significantly streamlined, reducing calibration time by 87.5\%, as only one out of eight reference pattern must be calibrated.
These findings highlight the potential of stimulation signal modulation to improve sensation quality and validate the viability of our predictive model for automating intensity calibration. This approach is an essential step toward delivering complex and naturalistic sensations in advanced haptic systems.
\end{abstract}

\begin{IEEEkeywords}
Electrotactile feedback, modulation, naturalness, perceived intensity, calibration, intensity prediction.
\end{IEEEkeywords}

\section{Introduction}
\IEEEPARstart{H}{aptic} feedback is an emerging field with the potential to transform various applications, including immersive experiences in virtual reality and gaming, enhancing the precision of teleoperation tasks, and restoring the sense of touch for prosthetics \cite{emami2024survey, shi2021self, zhao2020electrically, wang2022effective}. Among different haptic feedback techniques, electrotactile feedback stands out as a promising option. This method operates by applying electrical current signals to the skin surface to stimulate the nerves beneath \cite{saunders1983information, chouvardas2008tactile}. Compared to other haptic approaches, electrotactile feedback offers key advantages such as low cost, minimal power consumption, lightweight design, and high wearability \cite{chouvardas2008tactile, kourtesis2022electrotactile}. These features highlight the importance of further research and innovation to address its challenges and enhance its effectiveness.

One of the primary challenges limiting the widespread adoption of electrotactile feedback is achieving a high-quality sensation. Users frequently report that electrotactile stimulation induces tingling or buzzing sensations that feel unnatural \cite{tan2014neural, raspopovic2021sensory}. The naturalness of the sensation, defined as its resemblance to real touch, plays a crucial role in the overall quality of the experience. Researchers have attempted to understand and improve electrotactile sensations using multiple approaches, but these efforts remain somewhat fragmented.

Several research directions have been explored to enhance electrotactile feedback. One approach focuses on understanding the underlying mechanisms of touch perception in humans \cite{okorokova2018biomimetic}. Studies in this area investigate the properties and functions of skin and receptors, but the precise mechanisms by which electrical signals produce tactile sensations remain unknown. To bridge this gap, some researchers have developed simulation models to analyse the behaviour of skin and neurons under electrotactile stimulation \cite{araiza2019simulation}.

Another major research direction involves studying the effects of various electrotactile parameters including frequency, pulse width, amplitude, and electrode characteristics (e.g., size, shape, placement, material), on the perception of sensation, such as intensity, type, and quality \cite{zhou2022electrotactile}. For instance, studies have shown that stimulation frequency determines the type of mechanoreceptors activated, influencing the perceived sensation \cite{kajimoto1999tactile}. Similarly, other research has explored the impact of amplitude and pulse width on sensation intensity \cite{kaczmarek2017interaction, zhou2023exploring}. However, these studies often rely on simple tonic signals, such as mono-phasic or bi-phasic pulses with fixed frequency and amplitude, to systematically examine the effects of the parameters \cite{kaczmarek1991electrotactile, paredes2015impact}.

Recent studies have explored electrode shape modifications to improve sensation quality. For example, a study used elliptical electrodes instead of circular ones to better match the ergonomics of the finger \cite{isakovic2022impact}. Another study introduced rectangular and trapezoidal electrode designs of varying sizes to compare their effectiveness in electrical stimulation \cite{garenfeld2023novel}.

Research on waveform effects has also been conducted. One study evaluated three types of waveforms: a standard mono-phasic square wave, a square wave with time-varying pulse width, and a sine wave. The mono-phasic waveform produced clear tactile sensations, while the time-varying pulse width waveform generated more complex and nuanced sensations, potentially suitable for specific applications. The sine wave was found to produce a smoother but less distinct tactile sensation \cite{yang2019electrotactile}. These findings provide valuable insights for selecting appropriate waveforms to achieve various electrotactile feedback effects and enhance the quality of haptic feedback.

Despite these advancements, significant gaps remain. A key limitation of previous work is the reliance on simple tonic stimulation signals, which do not account for the temporal dynamics of real tactile experiences. These simplified stimulation signals often fail to replicate the mechanical patterns of natural touch and the corresponding neural responses, resulting in unnatural sensations. Although, as mentioned previously, one study explored a mono-phasic signal with time-varying pulse width, it did not investigate the effects of varying other parameters such as frequency and amplitude \cite{yang2019electrotactile}.

Temporal modulation of parameters is more common in neurological studies, where researchers have used invasive, implantable peripheral interfaces to stimulate the residual nerves of upper limb amputees. Different encoding strategies have been employed to modulate the stimulation parameters. For example, in one study, frequency and amplitude were linearly modulated \cite{valle2018comparison}. However, these studies primarily aimed to assess the impact of modulation on the location, extent, and intensity of perceived sensations in the phantom hand \cite{valle2018comparison, gholinezhad2023encoding, graczyk2016neural, graczyk2022frequency}.

Additionally, other studies have explored non-invasive methods for prosthetic applications. For example, in one study, researchers non-linearly co-modulated frequency and pulse width to improve the naturalness of electrical stimulation sensations in the foot \cite{bucciarelli2023multiparametric}. In another study, pulse width was modulated for electrodes placed on the upper arm, demonstrating that this modulation enabled the coverage of individual fingers or phalanges in a phantom hand. The researchers were able to present three distinct sensation levels (graded magnitudes) \cite{shin2018evoked}.

We believe that similar modulation strategies should be examined for other applications of electrotactile feedback beyond prosthetics. While these methods can serve as inspiration, independent studies are necessary due to differences in objectives, users, and stimulation locations. For example, the effects of frequency modulation on fingertip stimulation might differ due to variations in mechanoreceptor types and their frequency responses. As a result, identical frequency modulation strategies may yield different perceptual outcomes.

In this study, since achieving natural sensations requires complex signals that mimic the physics and mechanics of real touch interactions, we first propose a novel electrotactile system capable of generating complex, time-varying stimulation signals. We then design novel stimulation patterns in which amplitude, frequency, or both are linearly modulated to enhance the naturalness of sensation in a specific scenario. Linear modulation refers to the gradual change of electrical parameters over time in a controlled, linear manner. The specific tactile scenario we focus on is the sensation of pressing a large button with the index finger.

To validate our approach, we conduct a multistep experiment to collect data and evaluate both our electrotactile device and the designed stimulation patterns. Additionally, based on the collected data, we investigate the parameters influencing perceived intensity in modulated signals. Using these insights, we develop a predictive model that estimates the preferred intensity across different stimulation patterns based on a single known preference.

Our analysis reveals that temporal modulation of stimulation signals enhances sensation quality by improving its naturalness. Furthermore, we demonstrate that the energy of the optimal intensities selected by the participants correlates with the energy of different stimulation categories. This relationship enables us to propose an intensity prediction system that reliably estimates the preferred intensities for various stimulation patterns. This system is particularly beneficial for calibrating electrotactile feedback across multiple stimulation patterns for individual users. If these diverse stimulation patterns are to be implemented in real-world devices, manually determining the appropriate intensity for each pattern is a time-consuming process. Our predictive model streamlines this calibration process, reducing the time required for intensity adjustment across multiple stimulation categories.

The remainder of this paper is structured as follows. \cref{sec::methods} introduces the electrotactile device, signal modulation framework, and designed experiments. \cref{sec::analysis_results} explains the analysis of the collected data and discusses the experimental results. Finally, \cref{sec::discussion} outlines the main contributions, challenges, and future research directions.

\section{Methods}
\label{sec::methods}
\subsection{Custom Electrotactile Device}
\noindent The custom stimulation device had to meet various requirements for electrotactile feedback and the experiments that we planned to conduct.

A major feature of electrotactile feedback devices that distinguishes them from other haptic feedback methods, is their size, weight, wearability and price. So, for hardware and electronic design, we considered requirements such as being small, lightweight, and easy to wear and use.

The second consideration, with respect to both the hardware and the firmware, was the ability of the device to generate various bi-directional stimulation patterns and waveforms. We planned it to be as general as possible to keep our hands open for various experiments by enabling us to define our desired waveform to apply as stimulation.

For the safety requirements, the current amplitude is set to be limited to 3 mA in our hardware. Furthermore, a shutdown button is provided and available at all times during the experiments for the participants to use it in case of any noticeable discomfort or pain.

\begin{figure}[ht!]
\centering
\includegraphics[width=\linewidth]{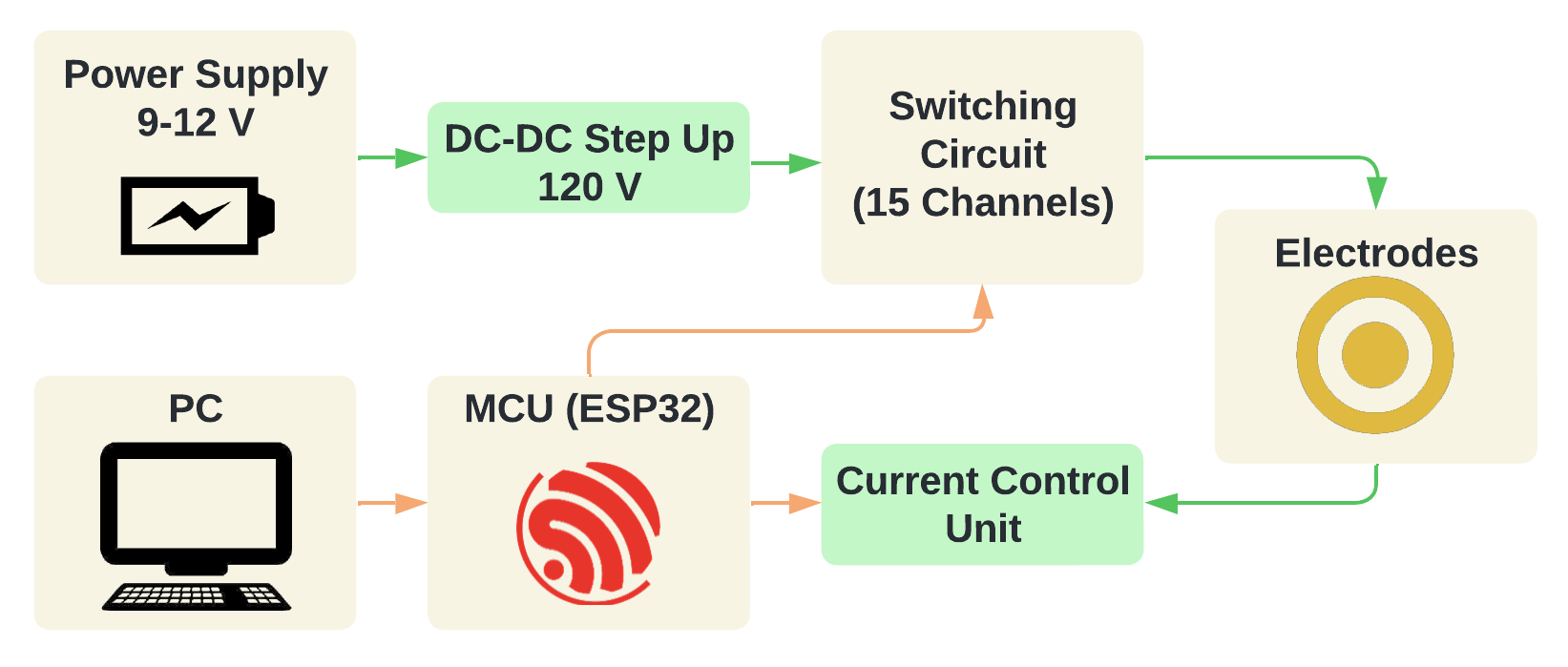}
\caption{Block diagram of the custom electrotactile device.}
\label{fig::electrotactile_device}
\end{figure}

The stimulation system is composed of three main parts including a voltage source, flexible electrodes to apply the signals to the hand, and a control circuit. \cref{fig::electrotactile_device} illustrates the block diagram of our custom electrotactile device and \cref{tab::stimulation_system_features} lists the features of it.

\begin{table}[b]
\caption[]{Capabilities of our Electrotactile Device}
\centering
\begin{tabular}{ll}
\hline 
\textbf{Parameters} & \textbf{Values} \\
\hline 
No. of Channels & 15 \\
Compliance Voltage & 120 V \\
Waveform & Mono-phasic, Bi-phasic (4 modes) \\
Frequency & $1-50 \mathrm{~kHz}, 1 \mathrm{~Hz}$ Resolution \\
Amplitude & $0-3 \mathrm{~mA}, 0.015$ to $0.035 \mathrm{~mA}$ Resolution \\
Pulse Width & $5-1000 \mathrm{us}, 1 \mathrm{us}$ Resolution \\
\hline
\end{tabular}
\label{tab::stimulation_system_features}
\end{table}

\subsubsection{Voltage Source}
The voltage source in the system consists of a power supply ranging from 9 to 12 V. Due to the high resistance of the electrode-skin interface, particularly for small dry electrodes, the compliance voltage required to deliver currents of up to 3 mA to the user's hand significantly exceeds 12 V \cite{george2020intensity}. To address this, a DC-DC step-up converter is employed, boosting the voltage to 120 V to meet the compliance voltage requirements, as is commonly done in the literature \cite{trout2023portable, yao2022encoding}. 

The input voltage range of 9 to 12 V also allows the use of lightweight batteries (weighing less than 150 g) that can provide up to 5 hours of operation. For a more compact design, smaller and lighter batteries (approximately 65 g) with 3 to 5 V can be utilised in conjunction with an additional voltage boost converter, offering up to 2 hours of operation, sufficient for a haptic feedback device.

\subsubsection{Electrodes}
To deliver stimulation in electrotactile devices, flexible electrodes have been developed using different materials and design patterns. The key requirement is flexibility, which ensures consistent and complete contact between the electrode pads and the curved surface of the skin, but good adhesion to maintain a reliable connection is also a key factor in an electrode \cite{huang2022recent, withana2018tacttoo}. 

Furthermore, electrodes in these systems utilise nanometre-thick metallic layers as conductive elements to deliver stimulation to the skin. Gold and silver are the most commonly used materials due to their flexibility, high conductivity, biocompatibility, and suitability for micro fabrication \cite{xu2016epidermal, withana2018tacttoo, yao2022encoding}.

In the initial phase of our work, we used electrode arrays designed on off-the-shelf flexible PCBs (Flex-PCBs). This simplified fabrication approach allows for the design of various electrode arrays and facilitates a range of experiments. The conductive layer of the electrode pads consists of a 55 $\mu$m thick copper layer, topped with a 75 nm thick gold layer. To secure the electrodes on the user's fingertips, we used medical tape, ensuring stability and preventing displacement.

In the experiments reported in this paper, we aimed to recreate the sensation of pressing a large button with the centre of the index finger. To achieve this, we utilised an electrode comprising two pads. As illustrated in \cref{fig::electrotactile_device} within the "Electrodes" box, the design features a small circular electrode pad (radius of 2 mm) encircled by a ring-shaped pad (radius of 5.3 mm and width of 2 mm). This configuration primarily elicited sensations beneath the small central pad, targeting the centre of the finger.

\subsubsection{Control Circuit}
Electrotactile devices generate current signals delivered to specific electrode pads, typically controlled by a microcontroller unit (MCU) which orchestrates the entire process. Various methods have been employed for the current generation, with two approaches being particularly prevalent: a) Howland Current Pump Topology: that utilises an operational amplifier and a balanced resistor bridge to create a voltage-controlled current source (used in \cite{trout2023portable}), b) Current Mirror Configuration: that employs transistors to replicate a reference current, ensuring consistent current delivery across multiple channels (used in \cite{yao2022encoding, yao2024fully}). In multi-channel devices, switching or multiplexing circuits are often implemented to distribute the current effectively across the electrode array.

We selected the ESP32 as the MCU for its affordability, availability, built-in modules, and wireless connectivity. For current generation and amplitude control, we opted for the improved Wilson current mirror configuration due to its simplicity, efficiency, and low price, avoiding the precise resistor matching required in the Howland topology and the limited availability of suitable integrated circuits (IC). The MCU regulates the command voltage of the current control unit via a digital-to-analogue converter (DAC).

To manage multiple channels, we used an analogue switch IC with a specific topology, assigning each channel one of four states: 'Source', 'Sink', 'Ground', 'No Connection'. Each channel corresponds to an electrode pad in the array, with the MCU controlling state transitions through the serial peripheral interface (SPI). This configuration enables bi-directional current flow between each pair or group of electrode pads, overcoming the limitations of mono-directional current and common-cathode designs found in devices such as those used in \cite{yao2022encoding, yao2024fully}. By dynamically adjusting these states, the MCU effectively regulates the frequency and pulse width of the stimulation current.

\subsubsection{Firmware}
The system is controlled by commands sent from a PC to the MCU, which generates signals by managing various system units. These commands specify waveform parameters, frequency, pulse width, amplitude, switch channel configurations, and duration of stimulation. The firmware also supports linear modulation of current amplitude and frequency throughout stimulation, enabling the recreation of mechanical pressure patterns in the stimulation signal.

A graphical user interface (GUI) facilitates defining stimulation patterns and generating corresponding commands, which are transmitted to the MCU to apply stimulation via flexible electrodes.

To enhance current amplitude configuration, instead of using a linear estimation of the mapping between the command voltage and the current that passes through the current mirror, we used a look-up table derived from an exact precise computer simulations, ensuring more accurate current control.

Currently, the device is not yet exactly wearable, as it relies on a wired PC connection for both power (only for MCU) and data transmission. However, its design and component selection allow for easy integration into a compact PCB. Wearability limitations can be addressed by using a DC-DC step-down converter for independent power supply and leveraging the ESP32’s built-in BLE and Wi-Fi for wireless data/command transfer.

With these modifications, the device can be miniaturised into a lightweight, low-cost wearable electrotactile feedback interface. A key advantage of our system is its ability to generate complex stimulation patterns. This is achieved through precise control over pulse timing, width, and amplitude, as well as independent state control of each electrode pad. Each pad can function as a current source, sink, ground, or remain disconnected, enabling dynamic spatial configurations for patterned stimulation signals.

\subsection{Designed Stimulation}
For designing a complex stimulation signal, we chose linear modulation for several reasons. First, it is a relatively simple form of modulation that can convey more information than tonic stimulation, while still allowing the effects of parameter changes to be easily measured and interpreted. This interpretability is essential, as understanding how changes in electrotactile feedback parameters influence sensation enables precise control and replication of these effects. Second, linear modulation is more cost-effective to implement compared to more complex stimulation schemes. Third, we believe that, in specific scenarios, modulating frequency, amplitude, and pulse width (individually or in combination) in a linear fashion can provide a more natural tactile experience and evoke a realistic neural response.

One such scenario is the sensation of pressing a button with a fingertip. As pressure starts, the fingertip first experiences a light touch. As the button is pressed further, the pressure intensifies, creating a deeper sensation in the finger. When the pressure is released, the sensation gradually decreases in a linear way. To mimic this real touch using an electrotactile stimulation, the user should feel the pressure start at a low intensity, gradually increase to a higher intensity, and maintain that intensity until the release phase begins, and gradually fade out after that. This process closely mimics a ramp-and-hold pattern, where the ramp sections represent a linear increase or decrease in intensity.

Incorporating this ramp-and-hold pattern into a stimulation signal can be achieved in several ways. First, by linearly modulating the amplitude of the electrotactile feedback signal. This approach replicates the mechanical pressure pattern, as the amplitude of electrical pulses directly correlates with the perceived intensity of the tactile feedback \cite{kourtesis2022electrotactile, akhtar2018controlling}. Consequently, linear modulation of amplitude mirrors the gradual intensity changes experienced during a button press.

Additionally, linear modulation of the frequency can further enhance the sensation. As pressure increases, the mechanoreceptors located in the deeper layers of the fingertip, which are responsible for sensing deep pressure, become activated. These mechanoreceptors are also more responsive to higher-frequency signals \cite{emami2024survey, purves2019neurosciences, goldstein2021sensation}. Higher frequencies not only stimulate deeper sensory neurons, but also reduce skin impedance, allowing the signal to penetrate more effectively \cite{xu2021effects}. Thus, increasing frequency with rising pressure helps simulate the experience of more intense or deeper pressure within the fingertip. However, it's important to note that higher frequencies may also increase the sensation of tingling, which could affect the overall tactile experience. 

By applying linear modulation to both amplitude and frequency, either separately or simultaneously, in a ramp-and-hold pattern, we aim to replicate the natural dynamics of pressing a button and activate sensory neurons in a way that feels more natural.
To test our hypothesis, we designed eight categories of stimulations which are grouped in four main categories:

\subsubsection{Tonic Stimulations}
These are fixed-frequency, fixed-amplitude signals throughout time, similar to the typical electrotactile feedback used in existing literature. We used two frequencies, 20 Hz and 100 Hz, to represent low and high frequencies, respectively. These values were chosen because they are commonly used in research and provide a good balance between clear differentiation and avoiding irritation \cite{kaczmarek2017interaction, paredes2015impact, manoharan2024characterization}. The amplitude will be chosen by the participants in a calibration process from 26 predefined values which will be explained in more details in the \cref{subsec::experiments}. The tonic category serves as a benchmark, allowing us to compare the modulated stimulations and evaluate whether modulation improves the sensory experience. These stimulations will be referred to as Tonic 20 Hz and Tonic 100 Hz throughout the rest of the paper.

\subsubsection{Linear Amplitude Modulation}
In this category, to recreate the ramp-and-hold pattern, the signal amplitude begins at a low value (e.g., 0.6 mA), increases linearly to a higher value (e.g., 0.9 mA) over 0.7 seconds (ramp-up time), holds at this peak for 1.6 seconds, and then decreases back to the initial value over 0.7 seconds (ramp-down time), totalling 3 seconds. The difference between the low and high amplitude values is fixed at 0.3 mA. Similar to tonic signals, the participants will select their optimal intensity from 26 predefined levels of high amplitude value, with the low value consistently 0.3 mA below the chosen high value. To allow for comparison, stimulations are tested at the same two frequencies of 20 Hz and 100 Hz. These stimulations will be referred to as Amp 20 Hz and Amp 100 Hz throughout the rest of the paper.

\subsubsection{Linear Frequency Modulation}
In this category, to replicate a ramp-and-hold pattern, the signal's frequency begins at 20 Hz, increases linearly to 100 Hz over 0.7 seconds, holds at 100 Hz for 1.6 seconds, and then decreases back to 20 Hz over 0.7 seconds. This cycle also lasts 3 seconds. For the second stimulation with frequency modulation, the same logic was repeated, but with a different frequency range, from 40 Hz to 170 Hz. This choice was made to assess whether the modulation pattern itself is more important than the specific frequency values. In these stimulations, the amplitudes are constant throughout the signal and will be chosen by the participants from 26 predefined levels based on their preferred intensity. These stimulations will be referred to as Freq 20-100 Hz and Freq 40-170 Hz throughout the rest of the paper.

\subsubsection{Combined Frequency and Amplitude Modulation} 
In this category, we simultaneously modulate both frequency and amplitude in a linear fashion, creating a more complex stimulation. The frequency values are the same as those used in the frequency modulation category, and the amplitude values will be changed similarly to linear modulation, the values of which are chosen by the participants from the same 26 predefined levels. These stimulations will be referred to as Both 20-100 Hz and Both 40-170 Hz throughout the rest of the paper.

\begin{figure}[!t]
\centerline{\includegraphics[width=0.9\linewidth]{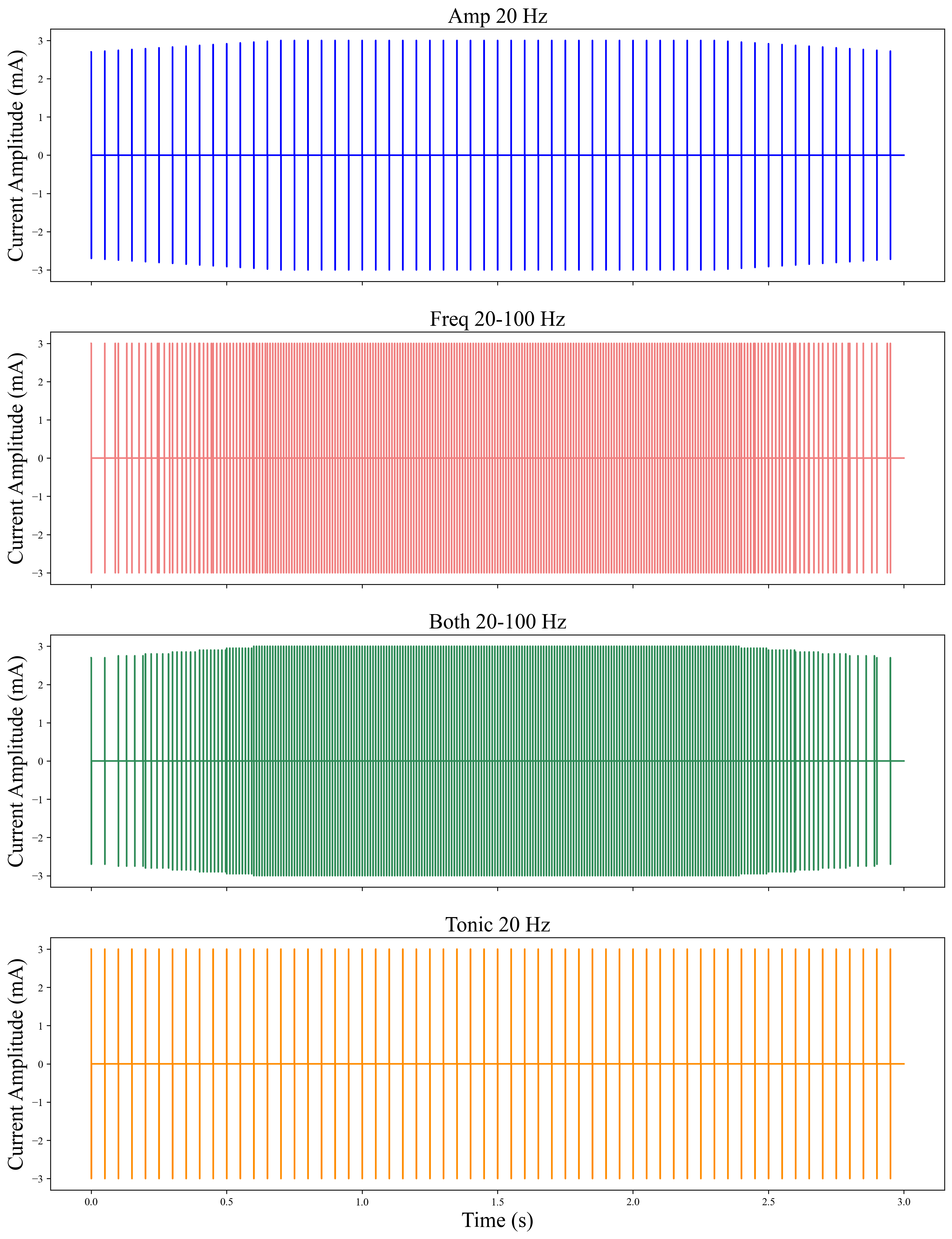}}
\caption{Different stimulation patterns. From top to down: the amplitude modulation in 20 Hz, the frequency modulation from 20 to 100 Hz, the simultaneous modulation by frequency changing from 20 to 100 Hz, and the tonic stimulation in 20 Hz}
\label{fig::modulated_signals}
\end{figure}

In this way, we designed eight categories of stimulations, all utilising bi-phasic pulses with equal positive and negative pulse widths of 300 $\mu$s, resulting in a total stimulation pulse width of 600 $\mu$s in each period of signal. Bi-phasic pulses were selected due to their widespread preference in electrotactile feedback applications, as they are easy to implement and interpret \cite{pamungkas2018overview}. Moreover, bi-phasic pulses offer significant advantages, including reduced risk of tissue damage and enhanced user comfort because of their charge balance, which prevents charge accumulation in the tissue. This minimises irritation and potential long-term adverse effects, making bi-phasic pulses particularly suitable for prolonged use and applications involving sensitive areas \cite{emami2024survey}. Furthermore, the use of square-shaped pulses facilitates faster depolarisation of nerve axons, thus improving the effectiveness and precision of stimulation \cite{pasluosta2018paradigms}. These features make bi-phasic square pulses an ideal choice for our stimulation design. \cref{fig::modulated_signals} presents a sample of a stimulation signal for each category. Each line represents a bi-phasic waveform.

\subsection{Experiments}
\label{subsec::experiments}
After designing the stimulations, we invited volunteers to participate in this study. Thirteen healthy participants (mean age: 31.6 years, age range: 24–53 years; 11 male, 2 female; 4 left-handed) volunteered for the experiments. The purpose of the study was explained to each participant, and informed consent was obtained prior to participation. This study was approved by the Ethical Committee of the University of Surrey, United Kingdom (FEPS 23–24 019 EGA).

Participants were seated comfortably in a chair in front of a computer screen displaying the GUI designed for the experiment. Experiment information and instructions were provided in written form several days before the experiment. On the day of the experiment, the instructions were also verbally explained to ensure clarity. Electrotactile stimulation was applied to the index finger of their dominant hand, which they were asked to rest on the table in front of them. The designed electrotactile device delivered the stimulation, while participants used their non-dominant hand to answer evaluation questions in the GUI.

Before attaching the electrodes, the skin on the volar side of the index finger was cleaned. The electrodes were placed on the fingertip and secured with a medical tape. Care was taken to ensure that the tape was snug but not so tight as to cause discomfort or a pulsating sensation due to restricted blood flow. To replicate realistic conditions, dry electrodes were used without the application of any conductive gel.

The electrode configuration for this experiment was fixed: the middle pad served as the source electrode, and the surrounding ring-shaped pad acted as the sink electrode. This configuration remained constant throughout all experiments.

The experiment consisted of two parts: a) Calibration to determine the optimal intensity of stimulation for each participant, b) Evaluation of the designed stimulations based on their perceived naturalness.

\subsubsection{Intensity Calibration}
In the calibration phase, participants were asked to adjust the intensity level of each stimulation to identify their optimal level of perceived intensity. To simplify the process and reduce the duration of the experiment, we fixed the positive and negative pulse widths at 300 $\mu s$ and intensity was solely varied by adjusting the amplitude of current.

We provide 26 predefined levels of intensity, starting from a low amplitude of 0.5 mA. Each subsequent level increased the amplitude by increments of 0.1 mA. Participants tested these levels to determine the intensity that was:
\begin{itemize}
    \item Strong enough to be easily detected.
    \item Not too strong to cause pain or irritation.
\end{itemize}

In addition to setting the optimal intensity for each stimulation category, participants were instructed to ensure that the perceived intensity felt consistent across categories, even if the actual amplitude levels differed. This step minimised the probable influence of intensity variations on the evaluation process and allowed us to focus solely on the sensation characteristics of the stimulations in the next phase of experiments.

Moreover, the data collected during this calibration phase provides an opportunity to analyse the relationship between stimulation intensity and signal parameters. While existing research has explored intensity for tonic stimulations, there is limited evaluation of intensity in the context of modulated, non-tonic signals.

The optimal intensity levels identified during this calibration phase were used for each participant in the subsequent evaluation phase.

\subsubsection{Sensation Evaluation}
In the evaluation phase, participants assessed the naturalness of the sensations evoked by the designed stimulations. Each stimulation category was repeated three times, and the order of the stimulations was randomised to prevent order effects and reduce error or bias. In total, participants evaluated 3 $\times$ 8 = 24 stimulations.

Participants were asked to rate the naturalness of each stimulation on a 5-point scale, where:

\begin{itemize}
    \item 0 represented a completely unnatural or uncomfortable sensation.
    \item 5 represented a completely natural sensation with no irritation.
\end{itemize}

We provided participants with a clear explanation of the expected sensation. Specifically, they were told that each stimulation was designed to mimic the sensation of pressing a large mechanical button during a course of 3 seconds. This explanation ensured that participants had a consistent understanding of the desired tactile experience when evaluating stimulations.

\section{Data Analysis and Results}
\label{sec::analysis_results}
\noindent Based on our research objectives and experimental design, we collected two primary datasets: a) Preferred intensity levels: The optimal stimulation intensities selected by each participant for each stimulation category. b) Naturalness scores: Ratings provided by participants for the perceived naturalness of each stimulation category. The analysis and results are outlined as follows:

\subsection{Intensity vs Energy of the Current signal}
Studies demonstrate that amplitude and pulse width strongly influence perceived intensity, and frequency also plays a role \cite{zhou2023exploring}. However, these findings are typically applied to tonic stimulations with a fixed amplitude and pulse width throughout the signal. For more complex signals with varying amplitude, pulse width, and frequency, new parameters may need to be identified to accurately model perceived intensity.

To investigate that, we hypothesised that signal energy could serve as a comprehensive metric. In other studies, the pulse energy and its relation to perceived intensity have been studied \cite{akhtar2018controlling}. However, signal energy, rather than the pulse energy, integrates the effects of varying amplitude, frequency, and pulse width, all together, potentially reflecting the overall characteristics of the stimulation signal in a specific time window. We define signal energy as the energy of the current signal, calculated using the standard formula for the energy of a signal, represented by the following equation:

\begin{equation}
E = \int_{0}^{T} |I(t)|^2\,dt
\end{equation}

Where $E$ represents the energy value, $I(t)$ denotes the current signal, and $T$ is the total duration of the signal (in this case, 3 seconds). In other words, signal energy is computed as the integral of squared current values ($I^2$) over the total stimulation duration of $T$. This energy metric encapsulates the combined effects of amplitude, frequency and pulse width. Amplitude directly influences the magnitude of $I(t)$, pulse width determines the duration of non-zero bi-phasic pulses, and frequency dictates the number of pulses within the signal duration. As a result, signal energy serves as a comprehensive representation of these key parameters, capturing their collective impact on stimulation characteristics.

To validate our hypothesis, we follow a two-step approach. First, we analyse the signal energy distribution across different categories to assess whether energy serves as a reliable metric. Then, we examine the energy levels of predefined stimulation intensities selected as the optimal perceived intensity, aiming to establish a relationship between signal energy and perceived intensity.

\begin{figure*}[t!]
\centering
    \subfloat[]{\includegraphics[width=0.45\textwidth]{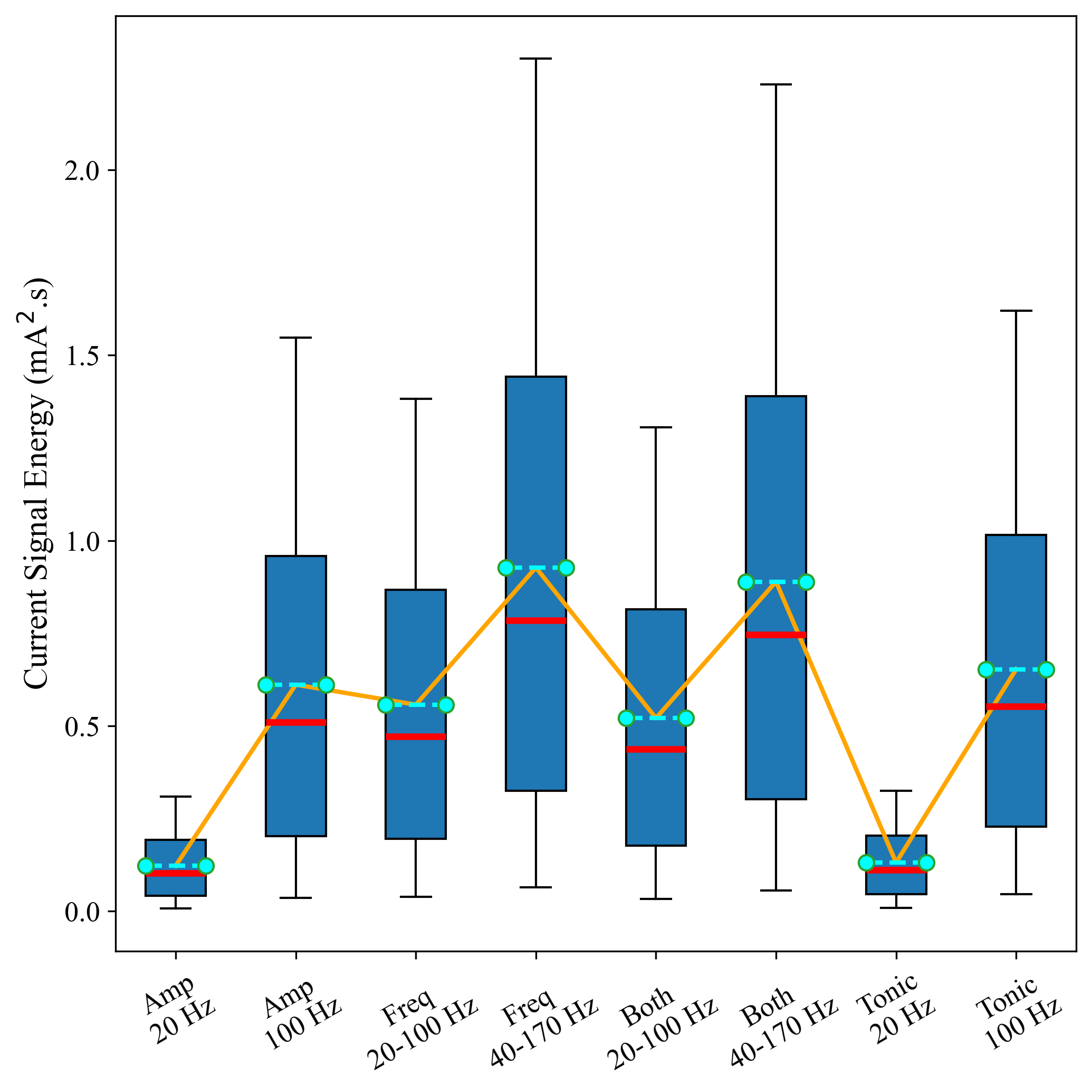} \label{fig::energy_of_signals::a}} 
    \subfloat[]{\includegraphics[width=0.45\textwidth]{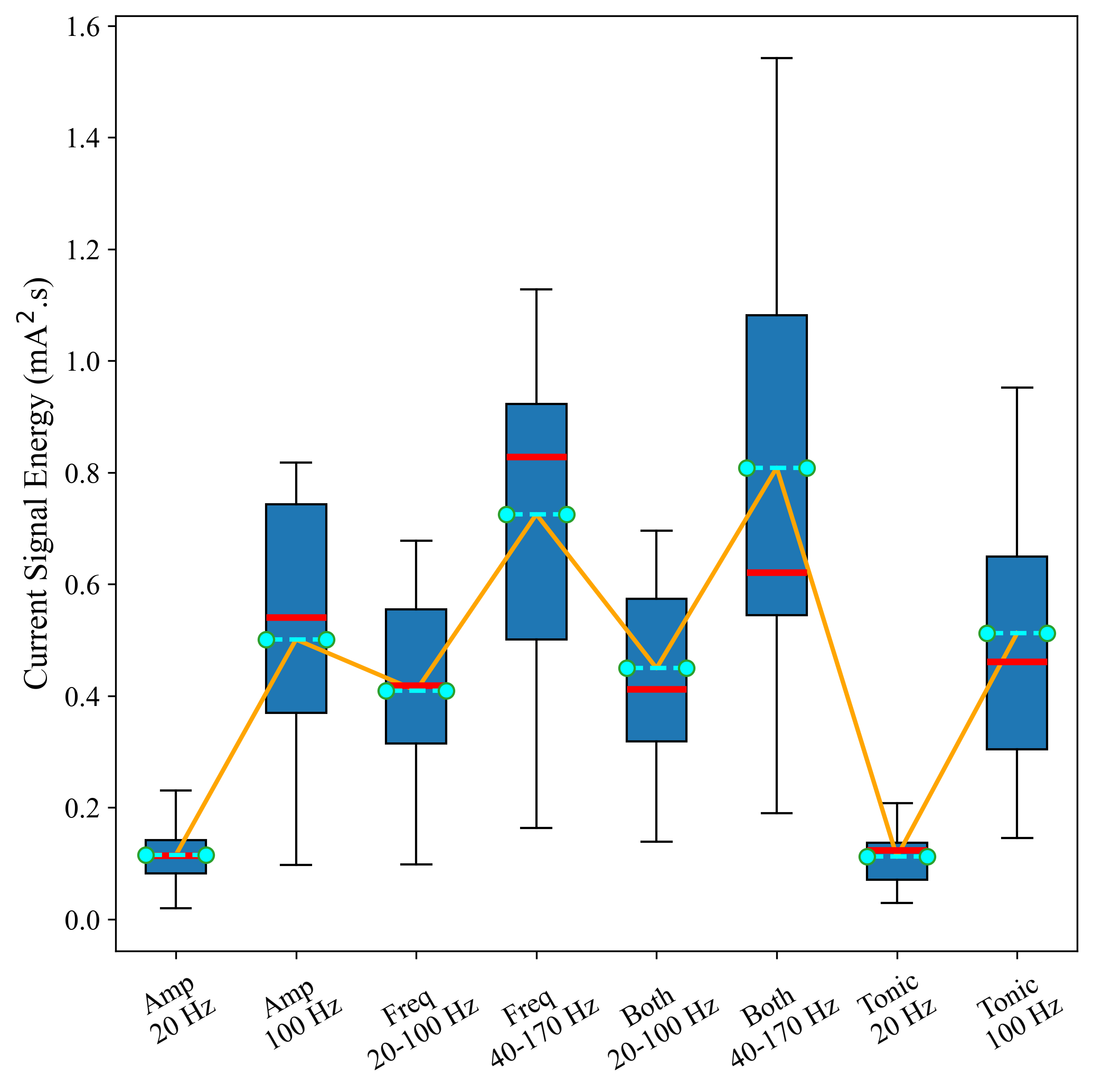} \label{fig::energy_of_signals::b}} 
    \caption{Comparison of the changes in the energy of a signal when pattern changes and energy of those patterns under same perceived intensity. (a) Energy of stimulation signals for different patterns in fixed-amplitude ranges. (b) Energy of selected stimulation signals by the participants for each pattern.}
    \label{fig::energy_of_signals}
\end{figure*}

\subsubsection{Signal Energy Distribution Across Categories}
We calculated the signal energy for each stimulation category across a range of amplitudes (0.5–3.0 mA, with 0.1 mA steps). \cref{fig::energy_of_signals::a} illustrates the energy distribution for all categories.
Key observations include:

\textbf{Frequency-Dependent Energy Ranges}: Categories with higher frequencies  ($\geq$ 100 Hz) exhibited significantly broader and higher energy ranges compared to those with lower frequencies (e.g., 20 Hz). This confirms that frequency substantially impacts the signal's energy profile.

\textbf{Amplitude Contributions}: Comparison of Amp 20 Hz and Tonic 20 Hz stimulations reveals that Amp 20 Hz exhibited slightly lower energy levels. This difference arises because the linear amplitude category incorporates a ramp-up and ramp-down phase, resulting in lower amplitudes at the signal's start and end compared to the constant amplitude of the Tonic category.

Thus, the signal energy distribution effectively captures both the frequency range and amplitude characteristics of the stimulation signals which makes it a comprehensive metrics.

\subsubsection{Relationship Between Signal Energy and Perceived Intensity}
\label{subsubsec::energy_vs_intensity}
We then analysed the relationship between signal energy and perceived intensity by examining the energy levels at the optimal intensities selected by participants for each category. \cref{fig::energy_of_signals::b} illustrates the distribution of energy across the selected stimulations.
Key findings include:

\textbf{Intensity Perception Independence}: Despite variations in signal energy across categories, participants reported the same perceived intensity at their selected optimal levels. This indicates that perceived intensity is not solely determined by signal energy, but is influenced by other factors. In a study, researchers observed that under the same perceived intensity, I$^2$T, where I represents pulse amplitude and T represents pulse width, is not constant but varies. They suggested that skin resistance is the reason, and showed that it changes linearly with I$^2$T \cite{akhtar2018controlling}. Our findings align with this observation, extending the analysis from individual pulses to the entire signal. This suggests that skin resistance may vary in response to changes in signal energy, indicating that overall signal energy could serve as a measure for monitoring skin resistance changes. This approach is particularly useful as it encompasses the influence of frequency, which is known to affect skin resistance \cite{xu2021effects}. Another key distinction in our study is the use of bi-phasic signals instead of mono-phasic ones, making our findings broader and potentially more generalisable.

\textbf{Energy Distribution Consistency}: The relationship between signal energy of different stimulation patterns at optimal intensities closely follows the overall energy distribution across those patterns. The orange line in both \cref{fig::energy_of_signals::a} and \cref{fig::energy_of_signals::b}, connecting the mean values, exhibits this similar trend.

This pattern of energy distribution across categories suggests the potential for a predictive relationship. Knowing the signal energy of a participant's preferred intensity for one stimulation pattern can allow for the estimation of signal energy of their preferred intensity for another pattern, ensuring consistent perceived intensity when using multiple stimulation patterns. This hypothesis forms the basis for developing a predictive model to automate and streamline calibration across stimulation categories.

\subsection{Intensity Predictive Model}
Our hypothesis posits that if the signal energy of the optimal intensity level is known for a specific reference stimulation (e.g., Tonic 100 Hz), it is possible to predict the signal energy corresponding to the same perceived intensity level for other stimulation categories. This prediction is grounded in the relationships between the energy distributions of different stimulation categories. In the first step, we tested our hypothesis by considering the mean values.

\subsubsection{Validation Using Mean of Energy Distribution}
The formula for this prediction can be expressed as:

\begin{equation}
E_i^{\text{pred}} = E_{\text{ref}}^{\text{selected}} \cdot 
\frac{\mathbb{E}[E_i]}{\mathbb{E}[E_{\text{ref}}]}
\end{equation}

Where $E_i^{\text{pred}}$ is the predicted energy of user's preferred intensity for stimulation category $i$; $E_{\text{ref}}^{\text{selected}}$ is the energy of the reference signal, calibrated to the user's preferred intensity level; $\mathbb{E}[E_i]$ is the mean energy of the distribution for stimulation category $i$; $\mathbb{E}[E_{\text{ref}}]$ is the mean energy of the distribution for the reference stimulation category.

To test this hypothesis, we calculated the mean signal energy for each category under consistent amplitude ranges (0.5–3.0 mA in our experiments). This condition ensures that the energy distributions are directly comparable since varying amplitude ranges could lead to differences in the energy range and invalidate meaningful comparisons of the means.

\begin{figure}[!t]
\centerline{\includegraphics[width=0.95\linewidth]{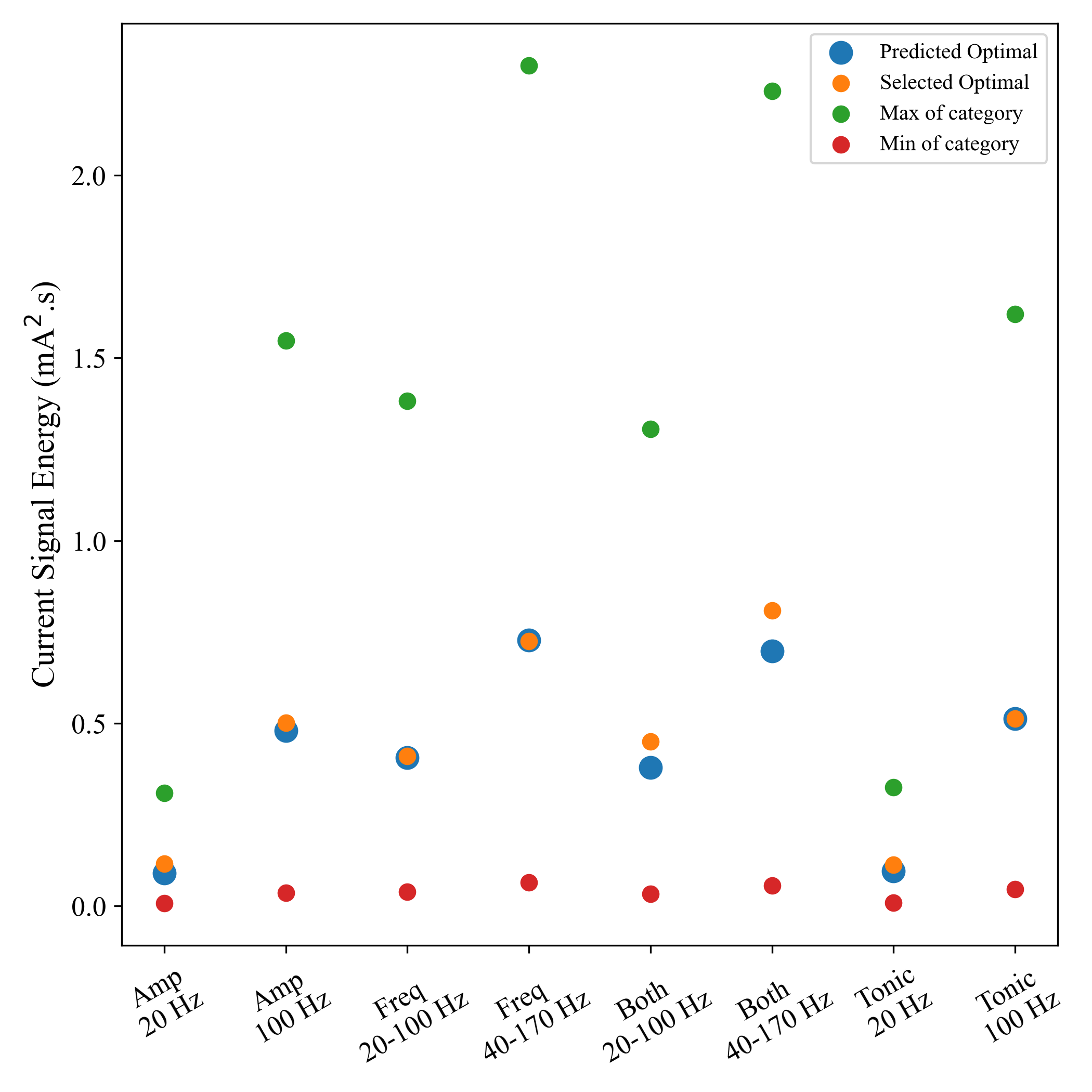}}
\caption{Comparison of predicted vs selected energy of desired intensity.}
\label{fig::predicted_vs_selected}
\end{figure}

Applying the above formula, we used the Tonic 100 Hz as the reference category and calculated the predicted optimal energy levels for all participants across the remaining categories. \cref{fig::predicted_vs_selected} presents the predicted energy values (blue dots) alongside the selected energy values (orange dots) for each category averaged though all participants. Additionally, the minimum and maximum energy levels present in the predefined levels (red and green dots, respectively) are plotted to contextualise the range of possible energy levels for each category. This approach allows us to evaluate how closely the predicted energy values align with the selected energy values within the range of possible energy levels. This comparison and the results showed that the predicted values closely matched the selected values across all categories. 

\paragraph{Prediction Evaluation}
To further validate the prediction model, we calculated the R$^2$ scores for predictions, grouping the results by stimulation categories and by participants. It should be noted that the reference stimulation (Tonic 100 Hz) has been removed, since we use its value as reference and its R$^2$ score will be 100\%. The R$^2$ score across participants are given in \cref{tab::R2score_participants} and the R$^2$ score across categories are given in \cref{tab::R2score_categories}.

The R$^2$ scores across participants yielded an average of 83.33\%, with some participants achieving scores exceeding 90\%. This high average confirms the effectiveness of the prediction model in capturing the relationship between energy distributions and desired intensity levels.

The R$^2$ scores across stimulation categories indicated that the predictions for some categories were less accurate than others. Notably, the two categories with the lowest R$^2$ scores were Amp 20 Hz and Tonic 20 Hz. These categories are characterised by their low frequencies, which suggests that dividing the stimulations to different groups based on their frequency range might improve the accuracy of the prediction.

\paragraph{Frequency-Based Grouping for Improved Prediction}

To test this argument, we divided the stimulation categories into two groups based on frequency:
\begin{itemize}
    \item High-frequency group: Stimulations with frequencies in range of 100 Hz or higher.
    \item Low-frequency group: Stimulations with frequency of 20 Hz.
\end{itemize}
For each group, we used a reference stimulation within the same frequency range:

\begin{itemize}
    \item Tonic 100 Hz was used as the reference for the high-frequency group.
    \item Tonic 20 Hz was used as the reference for the low-frequency group (applicable to Tonic 20 Hz itself and Amp 20 Hz).
\end{itemize}

Using this frequency-based grouping, the prediction accuracy improved significantly for Amp 20 Hz, as its R$^2$ improved nearly 50\% (achieved 87.671\% from 38.26\%). This highlights the importance of selecting an appropriate reference stimulation within the same frequency range.

Furthermore, by using the frequency-based grouping in the prediction, the overall mean R$^2$ score for participants increased to 84.27\%, with the reference stimulations (Tonic 100 Hz and Tonic 20 Hz) excluded from this calculation. While this categorisation improved the R$^2$ scores, the average increase across participants was marginal ($\sim$1\%). This modest gain should not be interpreted as an indication of methodological inefficacy but rather as a consequence of the specific conditions of our experiments. In our case, only two stimulation categories were categorised into a distinct low-frequency range, and for the remaining categories, the results were identical since they all used the same reference category (Tonic 100 Hz). Moreover, the frequency bands of our stimulations were relatively close to each other, which likely limited the differentiation between using one reference and using multiple references for each frequency band. However, in scenarios where stimulation frequencies span a wider range, such as including both low frequencies (e.g., 20 Hz) and very high frequencies (e.g., 500 Hz-2 kHz), frequency-based grouping could substantially enhance prediction accuracy.

\begin{table}[t]
\caption{R$^2$ Scores of Predictive Model for Each Participant}
\centering
\begin{minipage}[t]{0.45\linewidth} 
    \centering
    \begin{tabular}{ll}
    \hline 
    \textbf{PID} & \textbf{R$^2$ Score} \\
    \hline 
    P1  & 92.684 \\
    P2  & 56.555 \\
    P3  & 83.427 \\
    P4  & 89.913 \\
    P5  & 68.461 \\
    P6  & 95.341 \\
    P7  & 91.676 \\
    \hline
    \end{tabular}
\end{minipage}%
\begin{minipage}[t]{0.45\linewidth} 
    \centering
    \begin{tabular}{ll}
    \hline 
    \textbf{PID} & \textbf{R$^2$ Score} \\
    \hline 
    P8  & 87.482 \\
    P9  & 87.313 \\
    P10 & 67.482 \\
    P11 & 95.366 \\
    P12 & 74.940 \\
    P13 & 92.658 \\
    \hline
    \textbf{Average} & \textbf{83.331} \\
    \hline
    \end{tabular}
\end{minipage}
\label{tab::R2score_participants}
\end{table}

\begin{table}[t]
\caption[]{R$^2$ Scores of Predictive Model for Each Stimulation Pattern}
\centering
\begin{tabular}{ll}
\hline 
\textbf{Stimulation} & \textbf{R$^2$ Score} \\
\hline 
Amp 20 Hz & 38.26 \\
Amp 100 Hz & 85.549 \\
Freq 20-100 Hz & 85.527 \\
Freq 40-170 Hz & 90.116 \\
Both 20-100 Hz & 71.445 \\
Both 40-170 Hz & 84.041 \\
Tonic 20 Hz & 65.088 \\

\hline
\textbf{Average} & \textbf{74.289} \\
\hline
\end{tabular}
\label{tab::R2score_categories}
\end{table}

Furthermore, the enhancement observed with frequency-based grouping may be attributed to frequency-dependent variations in skin resistance. This suggests that, contrary to our discussion in \cref{subsubsec::energy_vs_intensity}, overall signal energy alone does not encompass all frequency-related effects, such as those on skin resistance. Our findings indicate that we cannot neglect the frequency range of a stimulation signal as it seems to be influencing peak skin resistance because, we showed that the signal energy of a stimulation pattern can be calibrated based on the signal energy of a reference stimulation within the same frequency range to achieve the same perceived intensity \cite{akhtar2018controlling}. Hence, if this result is further validated in future works, it may remove the need to measure skin resistance for each new pattern, impacting intensity calibration methods in future electrotactile devices.
 
\subsubsection{Validation Using Energy at Identical Amplitude Levels}
As previously discussed, using the mean energy for predictions requires defining the same amplitude range across different categories. To explore whether the energy values for identical amplitude levels could yield similar results, we extended our analysis. We compared the signal energy relationships between categories at matched amplitude levels rather than relying on their mean values. The revised formula can be expressed as:

\begin{equation}
E_i^{\text{pred}} = E_{\text{ref}}^{\text{selected}} \cdot 
\frac{E_i,_x}{E_{\text{ref}},_x}
\end{equation}

Where $E_i,_x$ is the energy of stimulation category $i$ at amplitude level of $x$, and $E_{\text{ref}},_x$ is the energy of reference stimulation at the amplitude level of $x$.

The results demonstrated that the relationship between categories remained consistent across these amplitude levels, confirming the general validity of the prediction formula. This finding suggests that even when amplitude ranges differ, as long as we select an identical amplitude level between categories, the prediction formula can still be applied effectively.

\subsubsection{Implications for Calibration}
The predicted energy values enable the calculation of intensity levels, as each intensity level corresponds to a specific energy value within each stimulation category. Once the intensity levels are determined, it is necessary to verify whether the device can produce the specified intensity levels within the constraints of its DAC resolution (or any other unit that controls current parameters). If certain intensity levels cannot be precisely generated, they can be approximated to the nearest possible value.

These findings highlight the feasibility of utilising a limited set of reference stimulation categories for the efficient calibration of electrotactile feedback systems. By calibrating the device for each individual using this small set of reference stimulations, the optimal intensity levels for all other stimulation patterns can be reliably predicted. This approach substantially reduces both the time and complexity involved in calibration while ensuring high accuracy and precision in delivering the desired tactile feedback with user calibrated intensity. In our case, with eight stimulation patterns, predicting the remaining seven based on a single reference category reduces calibration effort and time by 87.5\% (removing 7 out of 8), which is a substantial improvement.

\subsection{Perceived Naturalness of Stimulation Patterns}
\begin{figure}[t]
\centerline{\includegraphics[width=0.9\linewidth]{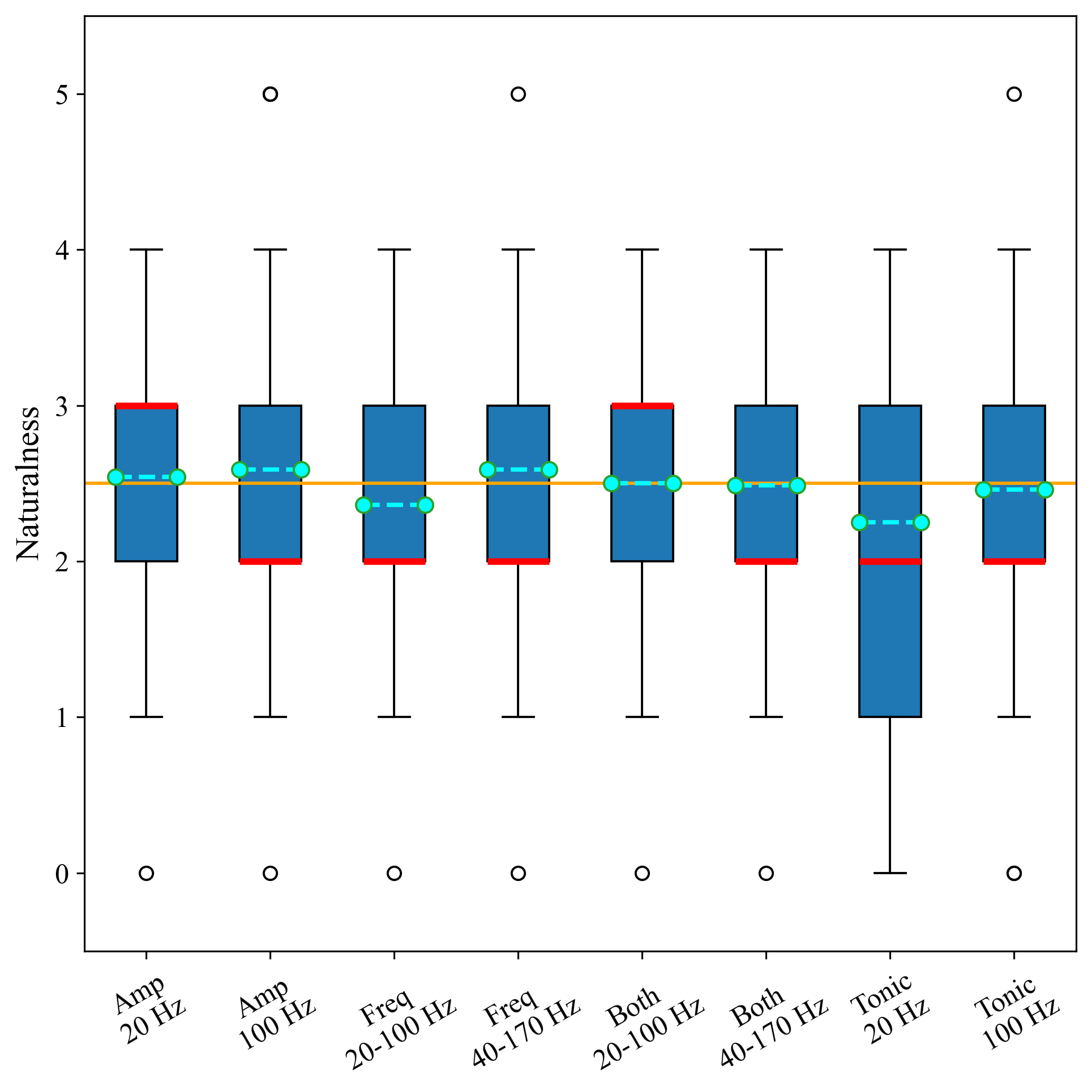}}
\caption{Mean (blue dashed line), median (red line), and interquartile ranges (boxes) of naturalness scores for each stimulation pattern.}
\label{fig::naturalness_result}
\end{figure}

The second part of our experiments focused on evaluating the naturalness of the evoked sensations. As described earlier, participants were asked to rate each stimulation on a scale from 0 to 5 (3 $\times$ 8 = 24 stimulations in total). The mean, median and interquartile ranges for these scores were then calculated. The \cref{fig::naturalness_result} presents these statistical values for each category, where the orange line indicates a score of 2.5 (the midpoint between 0 and 5), the blue dashed line represents the mean values, and the red line illustrates the median values.

\subsubsection{Overall Scores and Rankings}
Most naturalness scores fell between 2 and 3, indicating that the signals evoked sensations perceived by the participants as moderately natural. The ranked stimulations and their naturalness score are given in \cref{tab::natural_ranking}. Among the stimulations, Freq 40–170 Hz ranked the highest in naturalness, while Tonic 20 Hz ranked the lowest. The improvement in naturalness from the lowest to the highest score was approximately 6.8\%, highlighting the potential benefit of linear modulation. Notably, five out of the top eight rankings were assigned to modulated signals, underscoring the positive impact of modulation.

\begin{table}[!b]
\caption[]{Stimulations Ranking Based on Mean of Naturalness Scores}
\centering
\begin{tabular}{lll}
\hline 
\textbf{Rank} & 
\textbf{Stimulation} & \textbf{Mean Score} \\
\hline 
1 & Freq 40-170 Hz & 2.59 \\
2 & Amp 100 Hz & 2.59 \\
3 & Amp 20 Hz & 2.54 \\
4 & Both 20-100 Hz & 2.5 \\
5 & Both 40-170 Hz & 2.49 \\
6 & Tonic 100 Hz & 2.46 \\
7 & Freq 20-100 Hz & 2.36\\
8 & Tonic 20 Hz & 2.25 \\
\hline
\end{tabular}
\label{tab::natural_ranking}
\end{table}

\subsubsection{Impact of Modulation and Frequency}
To assess whether improvements in naturalness were due to modulation or frequency range alone, comparisons were made within the same frequency ranges:
\begin{itemize}
    \item For Amp 100 Hz (2nd rank) versus Tonic 100 Hz (6th rank), modulation improved naturalness by 2.6\%.
    \item For Amp 20 Hz versus Tonic 20 Hz, the improvement was more pronounced at 5.8\%.
\end{itemize}

These findings suggest that modulation enhanced the perceived naturalness of sensations. Additionally, the higher-frequency stimulations generally resulted in higher naturalness scores. This emphasises the importance of frequency selection, suggesting that optimising frequency values could further enhance naturalness beyond the levels observed in this experiment.

\subsubsection{Single vs. Combined Modulations}
Single modulations, such as Freq 40–170 Hz and the amplitude modulations, consistently outperformed combined modulations (Both 20–100 Hz and Both 40–170 Hz). This difference may be attributed to the increased informational burden of simultaneous modulations. However, it is worth noting that combined modulations still performed better than Freq 20–100 Hz, indicating the influence of frequency selection.

\subsubsection{Subjectivity and Median Scores}

The median scores revealed interesting subjective differences. While most categories had a median of 2, Amp 20 Hz and Both 20–100 Hz had a median of 3. This indicates that a larger proportion of participants rated these stimulations higher scores, although others gave very low scores, suggesting greater variability in perceptions.

This variability may stem from how participants perceived lower-frequency stimulations, which are often felt as vibrations rather than pressure. For some individuals, naturalness might align more with comfort, while for others, it is associated with the continuous sensation of pressure. These differences highlight the potential need for personalised stimulation parameters to optimise user experience.

\subsubsection{Maximum Scores and High-Frequency Potential}
Three categories, Freq 40–170 Hz, Amp 100 Hz, and Tonic 100 Hz, received a perfect score of 5 from some participants. All three involved high-frequency stimulations, demonstrating the potential of both modulation and higher frequencies in achieving more natural sensations. Across all metrics, Tonic 20 Hz consistently ranked as the least preferred stimulation for naturalness.

Our results differ from previous studies, where higher frequencies were often linked to stronger tingling sensations, leading participants to prefer lower frequencies \cite{manoharan2024characterization}. One possible reason for this difference is how preferences were assessed. In most studies, participants were simply asked what frequency they preferred, without a specific touch scenario. Since lower frequencies feel more comfortable and cause less tingling, they were usually preferred.

In our study, however, we asked participants to evaluate the stimulation in the context of pressing a button. This led them to find higher frequencies more natural, as lower frequencies are often associated with vibrations rather than continuous pressure \cite{mello2025towards}. This highlights the importance of designing future evaluation paradigms that are more application-specific, ensuring that user preferences are assessed within realistic usage contexts rather than abstract comparisons.

However, higher frequencies did not receive significantly higher scores across all participants. As mentioned before, this may be because individuals interpret "naturalness" differently, some prioritise realism, while others focus on comfort. These findings highlight the need for personalisation and a more comprehensive evaluation framework that accounts for both factors at the same time.

\section{Discussion}
\label{sec::discussion}
\subsection{Main Contributions}
\noindent In this study, we proposed a novel, precise, and versatile electrotactile feedback device. Using this device, we designed and implemented linearly modulated signals to enhance the naturalness of electrotactile sensations in specific scenarios, such as simulating the act of pressing a button. 

Through two-step experiments, with the participation of 13 subjects, we demonstrated several significant findings. First, we analysed the relationship between perceived intensity and signal energy under similar perceptual conditions. Our results showed a linear relationship between the energy of the signals preferred by participants as optimal intensity and the distribution of energy of the signal of that pattern for the fixed amplitude range, which is consistent with the findings of another study that investigated single-pulse stimulation. That study attributed this relationship to changes in skin impedance during stimulation and used it for real-time pulse calibration and providing constant perceived intensity during time \cite{akhtar2018controlling}. While our results may indicate a similar phenomenon, we focused on continuous signals instead of single pulses and did not directly measure skin impedance, leaving this aspect for future investigation. Nevertheless, the correlation between the signal energy of a reference stimulation pattern and an unknown pattern accurately predicted the user's preferred intensity level. By calibrating the desired intensity for various stimulation patterns based on a single user-selected signal energy, we significantly reduced the calibration time.

Second, our results confirmed that linear modulation of the signal improves the naturalness of the sensation compared to tonic stimulation, particularly in the scenario of mimicking a button press. The highest improvement in naturalness (6.8\%) was achieved using frequency modulation at higher frequency levels. This suggests that modulating temporal patterns of stimulation can convey richer and more realistic information, making the sensations feel more natural to users.
Moreover, the better performance observed with higher frequencies suggests that the impact of these parameters should be assessed using appropriate evaluation metrics and in scenarios that closely mimic real-world applications.

It is worth noting that the participants in our experiments represented a wide age range (24 to 53 years) and included both males and females, enhancing the robustness of our results with respect to the potential influence of age and gender.

\subsection{Challenges and Future Works}
One of the primary challenges in this study, as with many haptic feedback experiments, was the number of participants. While our sample size of 13 participants is relatively large compared to similar studies, a larger participant pool would further enhance the reliability and generalizability of our findings.

Our proposed intensity estimation model was tested on seven stimulation patterns, using one known intensity as a reference to predict the others. Although the results demonstrate strong predictive accuracy, future studies should evaluate this approach across a broader range of stimulation patterns and modulation techniques to confirm its robustness. Additionally, due to time constraints, the pulse width was fixed across all stimulations in our experiments. Since the pulse width plays a crucial role in perceived intensity, more research is needed to examine how varying the pulse width influences the relationship between signal energy and perceived intensity.

Although our findings indicate that linear modulation improves the naturalness of electrotactile feedback compared to tonic stimulation, it still remains a relatively simple modulation approach. The neural responses evoked by our modulated signals still do not fully replicate those produced by real-world tactile interactions, such as physically pressing a button. Although naturalness scores exceeded the average for electrotactile stimulation, they remain well below what would be considered a truly natural sensation. This highlights the need for further exploration of advanced modulation (e.g. non-linear) techniques and bio-inspired stimulations that better mimic real-world tactile experiences.

In this study, we assessed the naturalness of sensations using a numerical scoring system. This is an informative approach, but may not fully capture the complexity of user perception. Future work could benefit from a more comprehensive evaluation framework, such as detailed questionnaires or multidimensional rating scales, to gain deeper insights into how users perceive electrotactile feedback. It should be noted that this may significantly increase the duration of each experiment and thus limit the number of participants.

A key finding of this study is the effectiveness of temporal signal modulation in enhancing the naturalness of sensation. However, defining an appropriate temporal pattern requires determining the optimal time window for modulation. In our experiments, we used a fixed 3 seconds window to match the scenario of pressing a button. However, real-world touch interactions vary in duration, and using predefined time windows for each scenario is not practical for real-time applications. One approach to handling this in real-time is modifying pulses individually, but pulses individually cannot define a temporal pattern. Also, prediction of each pulse may impose a high computational cost, as the system would need to dynamically adjust each pulse based on real-time interactions.

To address this, we propose using stimulation units, defined as predefined signal patterns consisting of multiple pulses, not short like a single pulse and not as long as the whole tactile interaction, rather than controlling individual pulses. This stimulation units can have temporal patterns which probably is effective in quality of the sensation and are practical for real-time applications.

Also, we showed that the energy of a stimulation signal has a relation with perceived intensity, similar to findings in prior studies examining individual pulses \cite{akhtar2018controlling}. This suggests that the proposed stimulation units, rather than single pulses, could be used for intensity calibration and real-time skin impedance measurement, significantly reducing computational demands and improving energy efficiency in electrotactile devices.

An important open question is the optimal duration of stimulation units. A 100 milliseconds window may be too short for users to detect meaningful patterns, while a 3 seconds window may be unnecessarily long. Identifying the most effective time window for stimulation units is essential for maximising the benefits of temporal modulation while maintaining computational efficiency. Neurological research on how fast humans can detect changes in tactile sensations (temporal resolution of tactile stimuli) can help determining the optimal length for such stimulation units.

To determine the pattern of stimulation units based on real-time interactions, machine learning or artificial intelligence-driven models can be beneficial to dynamically adjust stimulation units based on user interactions. This represents a promising research direction that could significantly advance the realism and usability of electrotactile feedback systems. 

In this study, we employed a fixed electrode configuration. Future research should explore the impact of various electrode characteristics, such as pad shape and size, to further enhance electrotactile systems. Investigating electrodes with multiple pads and implementing spatial modulation within electrode arrays may improve the realism of electrotactile-evoked sensations.

\section{Acknowledgment}
\noindent The authors extend their heartfelt gratitude to everyone who contributed to the successful realisation of this work, as well as to the volunteers who generously participated in the experiment.

\IEEEtriggeratref{27}

\bibliographystyle{IEEEtran}
\bibliography{References.bib}

\end{document}